\documentstyle[12pt]{article}

\newtheorem{lemma}{Lemma}

\newtheorem{theorem}{Theorem}

\newtheorem{definition}{Definition}

\begin{document}

\title{
A Collection of Probabilistic Hidden-Variable Theorems and
Counterexamples\thanks{It is a pleasure to dedicate this article to
Giuliano Toraldo di Francia on the occasion of his 80th birthday.}}

\author{{\bf Patrick Suppes}\thanks{E-mail:
suppes@ockham.stanford.edu. To whom correspondence should be
addressed.} \and {\bf J. Acacio de Barros}\thanks{Permanent Address:
Physics Department, Federal University at Juiz de Fora, 36036-330 Juiz
de Fora, MG Brazil. E-mail: acacio@fisica.ufjf.br} \and {\bf Gary
Oas}\thanks{E-mail: oas@ockham.stanford.edu.} \\ 
{\it Ventura Hall,
Stanford University,} \\ {\it Stanford, California 94305-4115}}

\date{\today}

\maketitle

\newcounter{cms}

\setlength{\unitlength}{1mm}

\begin{abstract}

\vspace{1em}


\vspace{1em}

The purpose of this article is to formulate a number of probabilistic
hidden-variable theorems, to provide proofs in some cases, and
counterexamples to some conjectured relationships.  The first theorem
is the fundamental one.  It asserts the general equivalence of the
existence of a hidden variable and the existence of a joint
probability distribution of the observable quantities, whether finite
or continuous.

\end{abstract}


\newpage

The literature on hidden variables in quantum mechanics is now
enormous, and it may seem there is little that is new that can be
said.  Not everything in the present article is new, but several
things are.  We have tried to collect together a variety of results
that go beyond the standard Clauser-Horne-Shimony-Holt form of the Bell
inequalities for four observables.

First, we state, and sketch the proof, of the fundamental theorem of the
collection we consider: there is a factoring hidden variable for a
finite set of finite or continuous observables, i.e., random variables
in the language of probability theory, if and only if the observables
have a joint probability distribution.  The physically important
aspect of this theorem is that under very general conditions the
existence of a hidden variable can be reduced completely to the
relationship between the observables alone, namely, the problem of
determining whether or not they have a joint probability distribution
compatible with the given data, e.g., means, variances and
correlations of the observables.

We emphasize that although most of the literature is restricted to no
more than second-order moments such as covariances and correlations,
there is no necessity to make such a restriction.  It is in fact
violated in the fourth-order moment that arises in the well-known
Greenberger, Horne and Zeilinger \cite{GHZ} three- and four- particle
configurations providing new Gedanken experiments on hidden variables.
For our probabilistic proof of an abstract GHZ result, see Theorem 9.

As is familiar, Bell's results on hidden variables were mostly
restricted to $\pm 1$ observables, such as spin or polarization.  But
there is nothing essential about this restriction.  Our general
results cover any finite or continuous observables (Theorem 1).  We
also state a useful theorem (Theorem 7) on functions of random
variables, and give a partial corollary (Theorem 8) showing how such
general probabilistic results are implicit in the reduction of higher
spin cases to two-valued random variables in the physics literature. At
the end we give various results on hidden variables for Gaussian
observables and formulate as the final theorem a nonlinear inequality
that is necessary and sufficient for three Gaussian random variables
to have a joint distribution compatible with their given means,
variances and correlations.

\paragraph{Factorization.} In the literature on hidden variables, what
we call the principle of factorization is sometimes baptized as a
principle of locality.  The terminology is not really critical, but
the meaning is.  We have in mind a quite general principle for random
variables, continuous or discrete, which is the following. Let $ {\bf
X}_{1}, \ldots , {\bf X}_{n} $ be random variables, then a necessary
and sufficient condition that there is a random variable {\boldmath
$\lambda$}, which is intended to be the hidden variable, such that $
{\bf X}_{1} \ldots , {\bf X}_{n} $ are conditionally independent given
{\boldmath $\lambda$}, is that there exists a joint probability
distribution of ${\bf X}_{1}, \ldots , {\bf X}_{n} $, without
consideration of {\boldmath $\lambda$}. This is our first theorem,
which is the general fundamental theorem relating hidden variables and
joint probability distributions of observable random variables.

\begin{theorem} {\rm (}Suppes \& Zanotti \cite{SuppesZanotti81}, Holland \&
Rosenbaum, \cite{Holland}{\rm)} Let $n$ random variables $ X_{1}, \ldots ,
X_{n}$, finite or continous, be given. Then there exists a hidden
variable {\boldmath $\lambda$} such that there is a joint probability
distribution $F$ of $({\bf X}_1, \ldots ,{\bf X}_n,\mbox{\boldmath
$\lambda$})$ with the properties
\begin{description}
\item[(i)] $F( x_{1}, \ldots , x_{n} \mid  \lambda) =
P({\bf X}_1 \leq x_{1}, \ldots ,{\bf X}_n \leq x_{n} \mid \mbox{\boldmath
$\lambda$}=\lambda)$

\item[(ii)] Conditional independence holds, i.e., for all $x_{1},\ldots
,x_{n}, \lambda$, 
\[
F( x_{1}, \ldots , x_{n} \mid  \lambda) =
\prod_{j=1}^{n} F_{j}(x_{j}|\lambda),
\]
\end{description}
{\it if and only if} there is a joint probability distribution of ${\bf X}_1,
\ldots ,{\bf X}_n$. Moreover, {\boldmath $\lambda$} may be constructed
so as to be deterministic, i.e., the conditional variance given
{\boldmath $\lambda$} of each ${\bf X}_i$ is zero. 
\end{theorem}
To be completely explicit in the notation
\begin{equation}
F_j(x_j|\lambda) = P({\bf X}_j \leq x_j | \mbox{\boldmath
$\lambda$}=\lambda).
\end{equation}

\noindent{\it Idea of the proof.} Consider three $\pm 1$ random variables ${\bf
X}$, ${\bf Y}$ and ${\bf Z}$. There are $8$ possible joint outcomes
$(\pm 1, \pm 1, \pm 1)$. Let $p_{ijk}$ be the probability of outcome
$(i,j,k)$. Assign this probability to the value $\lambda_{ijk}$ of the
hidden variable {\boldmath $\lambda$} we construct. Then the probability of
the quadruple $(i,j,k,\lambda_{ijk})$ is just $p_{ijk}$ and the
conditional probabilities are deterministic, i.e., 
\[
P({\bf X} = i, {\bf Y} = j, {\bf Z} = k \mid \lambda_{ijk}) = 1,
\]
and factorization is immediate, i.e.,
\[
P({\bf X} = i, {\bf Y} = j, {\bf Z} = k \mid \lambda_{ijk}) = P({\bf
X} = i \mid \lambda_{ijk}) P ({\bf Y} = j \mid \lambda_{ijk} ) P ({\bf
Z} = k \mid \lambda_{ijk}).
\]

Extending this line of argument to the general case proves the joint
probability distribution of the observables is sufficient for
existence of the factoring hidden variable.  From the formulation of
Theorem $1$ necessity is obvious, since the joint distribution of
$({\bf X}_{1}, \ldots , {\bf X}_{n})$ is a marginal distribution of the
larger distribution $({\bf X}_{1} \ldots, {\bf X}_{n}, \mbox{\boldmath
$\lambda$}).$

It is obvious that the construction of {\boldmath $\lambda$} is purely
mathematical.  It has in itself no physical content. In fact, the
proof itself is very simple.  All the real mathematical difficulties
are to be found in giving workable criteria for observables to have a
joint probability distribution.  As we remark in more detail later, we
still do not have good criteria in the form of inequalities
for necessary and possibly sufficient conditions for a joint
distribution of three random variables with $n>2$ finite values, as in
higher spin cases.

When additional physical assumptions are imposed on the hidden
variable {\boldmath $\lambda$}, then the physical content of {\boldmath
$\lambda$} goes beyond the joint distribution of the observables.  A
simple example is embodied in the following theorem about two hidden
variables.  We impose an additional condition of symmetry on the
conditional expectations, and then a hidden variable exists only if the
correlation of the two observables is nonnegative, a strong
additional restriction on the joint distribution.  The proof of this
theorem is found in the article cited with its statement.

\begin{theorem} {\rm (}Suppes \& Zanotti \cite{SuppesZanotti80}{\rm )} Let ${\bf X}$
and ${\bf Y}$ be two-valued random variables, for definiteness, with
possible values $1$ and $-1$, and with positive variances, i.e.,
$\sigma({\bf X})$, $\sigma({\bf Y}) > 0$. In addition, let ${\bf X}$ and
${\bf Y}$ be exchangeable, i.e.,
\[
P({\bf X} = 1, {\bf Y} = -1 ) = P({\bf X} = -1, {\bf Y} = 1).
\]
Then a necessary and sufficient condition that there exist a hidden
variable {\boldmath $\lambda$} such that 

\[
E ({\bf XY} \mid \mbox{\boldmath $\lambda$} = \lambda) = E ({\bf X}
\mid \mbox{\boldmath $\lambda$} = \lambda) E ({\bf Y} \mid \mbox{\boldmath $\lambda$} = \lambda)
\]
and
\[
E({\bf X} \mid \mbox{\boldmath $\lambda$} = \lambda) = E ({\bf Y} \mid
\mbox{\boldmath $\lambda$}
= \lambda)
\]
for every value $\lambda$ (except possibly on a set of measure zero)
is that the correlation of ${\bf X}$ and ${\bf Y}$ be nonnegative.
\end{theorem}
The informal statement of Theorems 1 and 2, which we call the {\em
Factorization Theorems}, is that the necessary and sufficient
condition for the existence of a factorizing hidden variable
{\boldmath $\lambda$} is just the existence of a joint probability
distribution of the given random variables ${\bf X}_i$.

Often, in physics, as in the present paper, we are interested only in
the means, variances and covariances -- what is called the
second-order probability theory, because we consider only second-order
moments. We say that a hidden variable {\boldmath $\lambda$} satisfies
the {\em Second-Order Factorization Condition} with respect to the
random variables ${\bf X}_1, \ldots ,{\bf X}_n$ whose two first
moments exist if and only if

\vspace{.2in}
(a) $E({\bf X}_1 \cdots {\bf X}_n|\mbox{\boldmath $\lambda$})
= E({\bf X}_1|\mbox{\boldmath $\lambda$})\cdots E({\bf
X}_n|\mbox{\boldmath $\lambda$})$,

\vspace{.2in}
(b) $E({\bf X}_1^2 \cdots {\bf X}_n^2|\mbox{\boldmath $\lambda$})
= E({\bf X}_1^2|\mbox{\boldmath $\lambda$})\cdots E({\bf
X}_n^2|\mbox{\boldmath $\lambda$})$.

\vspace{.2in}

We then have as an immediate consequence of Theorem 1 the following.
\begin{theorem}
Let $n$ random variables discrete or continuous be given. If there is a
joint probability distribution of ${\bf X}_1, \ldots ,{\bf X}_n$, then
there is a deterministic hidden variable {\boldmath $\lambda$} such
that {\boldmath $\lambda$} satisfies the Second-Order Factorization
Condition with respect to ${\bf X}_1, \ldots ,{\bf X}_n$.
\end{theorem}

\paragraph{Locality.}
The next systematic concept we want to discuss is locality.  We mean
by locality what we think John Bell meant by locality in the following
quotation from his well-known 1966 paper \cite{Bell66}.
\begin{quotation}
It is the requirement of locality, or more precisely that the result
of a measurement on one system be unaffected by operations on a
distant system with which it has interacted in the past, that creates
the essential difficulty. ... The vital assumption is that the result
$B$ for particle 2 does not depend on the setting {\bf a}, of the
magnet for particle 1, nor $A$ on {\bf b}.
\end{quotation}
Although Theorems 1 and 2 are stated at an abstract level without any
reference to space-time or other physical considerations, there is an
implicit hypothesis of locality in their statements. To make the
locality hypothesis explicit, we need to use additional concepts. For
each random variable ${\bf X}_i$, we introduce a random vector ${\bf M}_i$ of
parameters for the local apparatus (in space-time) used to measure the
values of random variable ${\bf X}_i$.
\begin{definition}[Locality Condition I] 
\[
E({\bf X}_i^k|{\bf M}_i, {\bf M}_j, \mbox{\boldmath $\lambda$}) = E({\bf
X}_i^k|{\bf M}_i, \mbox{\boldmath $\lambda$}),
\]
where $k=1,2$, corresponding to the first two moments of ${\bf X}_i$,
$i\neq j$, and $1 \leq i,j\leq n$. 
\end{definition}
Note that we consider only ${\bf M}_j$ on the supposition that in a given
experimental run, only the correlation of ${\bf X}_i$ with ${\bf X}_j$
is being studied.  Extension to more variables, as considered in
Theorem 7, is obvious. In many
experiments the direction of the measuring apparatus is the most
important parameter that is a component of ${\bf M}_i$.

\begin{definition}[Locality Condition II: Noncontexuality]
The distribution of {\boldmath $\lambda$} is independent of the
parameter values ${\bf M}_i$ and ${\bf M}_j$, i.e., for all functions $g$ for which
the expectation $E(g(\mbox{\boldmath $\lambda$}))$ and
$E(g(\mbox{\boldmath $\lambda$})|{\bf M}_i, {\bf M}_j)$ are finite,
\[
E(g(\mbox{\boldmath $\lambda$})) = E(g(\mbox{\boldmath
$\lambda$})|{\bf M}_i,
{\bf M}_j) .
\]
\end{definition}
Here we follow \cite{Suppes-76}.  In terms of Theorem 3, locality in
the sense of Condition I is required to satisfy the hypothesis of a
fixed mean and variance for each ${\bf X}_i$. If experimental
observation of ${\bf X}_i$ when coupled with ${\bf X}_j$ was different
from what was observed when coupled with ${\bf X}_{j'}$, then the
hypothesis of constant means and variances would be violated. The
restriction of Locality Condition II must be satisfied in the
construction of {\boldmath $\lambda$} and it is easy to check that it
is. This is often called, as indicated, Noncontexuality.

We embody these remarks in Theorem 4.
\begin{theorem}
Let $n$ random variables ${\bf X}_1, \ldots ,{\bf X}_n$ be given
satisfying the hypothesis of Theorem 2. Let $M_i$ be the vector of
local parameters for measuring ${\bf X}_i$, and let each ${\bf X}_i$
satisfy Locality Condition I. Then there is a hidden variable
{\boldmath $\lambda$} satisfying Locality Condition II and the
Second-Order Factorization Condition if there is a joint
probability distribution of ${\bf X}_1, \ldots ,{\bf X}_n$.
\end{theorem}

\paragraph{Inequalities for three random variables.} The next theorem states
two conditions equivalent to an inequality condition given in
\cite{SuppesZanotti81} for three random variables having just two
values.
\begin{theorem}
Let three random variables ${\bf X, Y}$ and ${\bf Z}$ be given with
values $\pm 1$ satisfying the symmetry condition $E({\bf X}) = E
({\bf Y}) = E({\bf Z}) = 0$ and with covariances $E({\bf XY}), E({\bf YZ})$
and $E({\bf XZ})$ given. Then the following three conditions are
equivalent.
\begin{description}
\item[(i)] There is a hidden variable {\boldmath $\lambda$} with
respect to ${\bf X}$, ${\bf Y}$, and ${\bf Z}$ satisfying Locality
Condition II and the Second-Order Factorization Condition holds.

\item[(ii)] There is a joint probability distribution of the random
variables ${\bf X}$, ${\bf Y}$, and ${\bf Z}$ compatible
with the given means and expectations.

\item[(iii)] The random variables ${\bf X}$, ${\bf Y}$ and
${\bf Z}$ satisfy the following inequalities.

\[ -1 \leq E({\bf XY}) + E({\bf YZ}) + E({\bf XZ}) \leq 1 + 2
\mbox{{\rm Min}} (E({\bf XY}), E({\bf YZ}), E({\bf XZ})).  \]
\end{description} \end{theorem} There are several remarks to be made
about this theorem, especially the inequalities given in (iii). For
discussion we introduce the standard correlation, and its standard
notation, for two random variables ${\bf X}$ and ${\bf Y}$ whose
variances are not zero:
\[ \rho(X,Y) = \frac{E({\bf XY}) - E({\bf X})E({\bf Y})}{\sigma({\bf
X}) \sigma({\bf Y})}, \]
where $\sigma(X)$, $\sigma(Y)$ are the standard deviations of ${\bf
X}$ and ${\bf Y}$, i.e., the square roots of the variances: 
\[ \sigma(X) = \sqrt{\sigma^{2}({\bf X})} \]
and
\[ \sigma^{2}({\bf X}) = \mbox{Var} ({\bf X}) = E({\bf X}^{2}) -
E({\bf X})^{2}. \]
First, the explicit correlation notation $\rho({\bf X}, {\bf Y})$ is
not standard in physics, but is necessary here for comparing various
theorems. The notation adopted throughout this article conforms fairly
closely to what is standard in mathematical statistics.

Physicists use less general notation, because they often assume
certain symmetry conditions are satisfied, e.g., $E({\bf X}) = E({\bf
Y}) = E({\bf Z}) = 0.$ To make these relations explicit, keeping in mind the
earlier definition of $\rho({\bf X}, {\bf Y}),$ we have: 
\begin{description}
\item[(i)] Covariance of ${\bf X}$ and ${\bf Y} = \mbox{Cov}({\bf
X},{\bf Y}) = E({\bf XY}) -
E({\bf X})E({\bf Y})$,

\item[(ii)]If $E({\bf X}) = E({\bf Y}) = 0$, then, clearly,
\[ \mbox{Cov}({\bf X},{\bf Y}) = E({\bf XY}) \]

\item[(iii)] If ${\bf X}$ and ${\bf Y}$ are random variables whose
only values are $\pm 1$ and $E({\bf X}) = E({\bf Y}) = 0$, then

\[ \mbox{Var}({\bf X}) = \mbox{Var}({\bf Y}) = 1, \]

\item[(iv)] If hypothesis of (iii) is satisfied \[ \rho({\bf X},{\bf
Y}) = E({\bf XY}), \] 
\end{description}
which is why in the physics literature $E({\bf XY})$, with or without
a comma between ${\bf X}$ and ${\bf Y}$, so commonly occurs. The
statistical terminology for $E({\bf XY})$ is bivariate product moment
$\mu_{11}$, which we shall often simply call the bivariate product
moment, without further notation.

Note that with the special symmetry conditions that $E({\bf X}) =
E({\bf Y}) = E({\bf Z}) = 0$, the inequalities (iii) of Theorem 5 for 
$\pm$1 random variables  can be written
\begin{equation}
 -1 \leq \rho ({\bf X},{\bf Y}) + \rho ({\bf Y}, {\bf Z}) + \rho
({\bf X}, {\bf Z}) \leq 1 + 2 \mbox{Min}(\rho({\bf X}, {\bf Y}), \rho ({\bf
X}, {\bf Z}), \rho ({\bf Y}, {\bf Z})).
\end{equation}

\paragraph{Three Counterexamples.}To show how special (iii) of Theorem
5, or the equivalent (2) written in terms of correlation, is, because
of the strong symmetry assumptions, we now
give three different examples that do not satisfy these inequalities.
The first is for $\pm 1$ random variables that do not have
expectations equal to zero. For this case neither the correlations nor
covariances have linear inequalities, only the moments $E({\bf
XY})$. The second case is for random variables with values $-1, 0, 1$
and zero expectations. An example is given which is satisfied by the
covariances but not the correlations.  The third case is for random
variables with values $-2, 0, 2$ and zero expectations. The
inequalities of (iii) are not satisfied by the covariances, which in
this case are equal to the expectations $E({\bf XY})$.

First, for the general case of $\pm 1$ random variables we have 
\[ E({\bf X}) = x_{0},\, E({\bf Y}) = y_{0},\, E({\bf Z}) = z_{0} \]
and
\[ -1 < x_{0}, y_{0}, z_{0} < 1, \]
and it is straightforward to derive the analogue of (iii) of Theorem 5
for the bivariate product moments, as well as the corresponding
correlations, but the expressions are more complicated for the
correlations. We only give part of the details here. We generalize on
the derivation given in \cite{SuppesZanotti81}.  We need to consider
in detail the eight probabilities $p_{ijk}$ for $i,j,k = \pm 1.$ When
referring to the marginals we use a dot for the missing random
variable. For example,

\[
\begin{array}{lll}
p_{11\cdot} & = & P({\bf X} = 1, {\bf Y} = 1) \\
p_{0 \cdot 1} & = & P({\bf X} = -1, {\bf Z} = 1)
\end{array}
\]
(For ease of typography we use $0$ rather than $-1$ as a
subscript.) 

We note immediately the following equations:

\[ E({\bf XY}) = p_{11 \cdot} - p_{10 \cdot} - p_{01 \cdot} + p_{00
\cdot} \]

\[ p_{10 \cdot} + p_{01 \cdot} = \frac{1 - E({\bf XY})}{2} \]
and correspondingly,

\[ p_{\cdot 10} + p_{\cdot 01} = \frac{1 - E({\bf YZ})}{2} \]

\[ p_{1 \cdot 0} + p_{0 \cdot 1} = \frac{1 - E({\bf XZ})}{2} \]

\[ p_{1 \cdot \cdot} = p_{10 \cdot} + p_{11 \cdot} = \frac{x_{0} + 1}{2} \]

\[ p_{\cdot 1 \cdot} = p_{11 \cdot} + p_{01 \cdot} = \frac{y_{0}+ 1}{2} \]

\[ p_{\cdot \cdot 1} = p_{\cdot 11} + p_{\cdot 01} = \frac{z_{0} +
1}{2} \]
From these equations we easily derive 
\[ p_{10 \cdot} = \frac{x_{0} - y_{0}}{4} + \frac{1 - E({\bf XY})}{4}
\]

\[ p_{11 \cdot} = \frac{1}{4} + \frac{x_{0} + y_{0} + E({\bf XY})}{4},
\]
and similar expressions for $p_{1 \cdot 0}$, $p_{1 \cdot 1}$, etc.
Using these equations, we may then derive 

\[ p_{110} = \frac{1}{4} + \frac{x_{0} + y_{0} + E({\bf XY})}{4} -
p_{111} \]

\[ p_{101} = \frac{1}{4} + \frac{x_{0} + z_{0} + E({\bf XZ})}{4} -
p_{111} \]

\[ p_{011} = \frac{1}{4} + \frac{y_{0} + z_{0} + E({\bf YZ})}{4} -
p_{111} \] 

\[ p_{100} = p_{111} - \frac{y_{0} + z_{0}}{4} - \frac{E({\bf XY})}{4}
- \frac{E({\bf XZ})}{4} \]

\[ p_{010} = p_{111} - \frac{x_{0} + z_{0}}{4} - \frac{E({\bf XY})}{4}
- \frac{E({\bf YZ})}{4} \]

\[ p_{001} = p_{111} - \frac{x_{0} + y_{0}}{4} - \frac{E({\bf XZ})}{4}
- \frac{E({\bf YZ})}{4} \]

\[ p_{000} = \frac{1}{4} - \frac{x_{0} + z_{0}}{4} + \frac{E({\bf
XZ})}{4} - \frac{y_{0} + z_{0}}{4} + \frac{E({\bf YZ})}{4} -
\frac{x_{0} + y_{0}}{4} + \frac{E({\bf XY})}{4} - p_{111} \]
so
\[ 1 + E({\bf XY}) + E({\bf YZ}) + E({\bf XZ}) - 2(x_{0} + y_{0} +
z_{0} ) \geq 4p_{111} \]
And as a generalization of the left-hand inequality of (iii) of
Theorem 5, we then have
\begin{equation}
E({\bf XY}) + E({\bf YZ}) + E({\bf XZ}) - 2(x_{0} + y_{0} + z_{0})
\geq -1.
\end{equation}
This result is much simpler than the corresponding one for
correlation. We have at once 
\[ \rho ({\bf X}, {\bf Y}) = \frac{E({\bf XY}) -
x_{0}y_{0}}{\sqrt{1 - x_{0}^{2}} \sqrt{1 - y_{0}^{2}}}, \]
and so
\[ E({\bf XY}) = \sqrt{1 - x_{0}^{2}} \sqrt{1 - y_{0}^{2}} \rho ({\bf
X}, {\bf Y}) + x_{0}y_{0}. \]
Substituting the right-hand side for $E({\bf XY})$, and the
corresponding expressions for $E({\bf YZ})$ and $E({\bf XZ})$ yields a
rather complicated inequality in terms of correlation, which we shall
not write out here.

The next remark is that (iii) is not necessary for the correlations of
three-valued random variables with expectations equal to zero. Let the
three values be $1,0, -1$. Here is a counterexample where each of the
three correlations is $- \frac{1}{2}$, and thus with a sum equal to $-
\frac{3}{2}$, violating (2).

There is a joint probability distribution with the following values.
Let $p(x,y,z)$ be the probability of a given triple of values, e.g.,
$(1, -1, 0)$.  Then, of course, we must have for all $x,y \mbox { and
} z$
\[p(x,y,z) \geq 0 \mbox {  and  } \sum_{x,y,z} p(x,y,z) = 1, \]
where $x,y \mbox {  and  } z$ each have the three values $1,0, -1$. 
So, let
\[p(-1,0,1)\!=\!p(1,-1,0)\!=\!p(0,1,-1)\!=\!p(1,0,-1)\!=\!p(-1,1,0)\!=\!p
(0,-1,1)\!=\!\frac{1}{6} \]
and the other 21 $p(x,y,z) = 0$. Then it is easy to show that in
this model $ E({\bf X})\!=\!E( {\bf Y})\!=\!E ({\bf Z})\!=\!0, \mbox{Var} ({\bf
X})\!=\!\mbox{Var} ({\bf Y})\!=\!\mbox{Var}({\bf Z})\!=\!\frac{2}{3},
\mbox{and Cov}({\bf XY})\!=\!\mbox{Cov}({\bf YZ})\!=\!\mbox{Cov}({\bf XZ})\!=\!- \frac{1}{3}$, so that the correlations are
\[\rho ({\bf X,Y}) = \rho ({\bf Y,Z}) = \rho ({\bf X,Z}) = - \frac{1}{2}. \]
Note that in the example just given the covariances for the
three-valued random variables, with the joint distribution as stated,
do satisfy (iii) of Theorem 5. 

For the third promised case, it is easy to construct a counterexample
for covariances of three-valued random variables with values -2, 0, 2
and expectations zero. We use the same distribution for these new
values: $p(-2,0,2) = p(2,-2,0) = p(0,2,-2) = p(2,0,-2) = p(-2,2,0) =
p(0,-2,2) = \frac{1}{6}.$ It is easy to see at once that
\[ \mbox{Cov}({\bf X}, {\bf Y}) = \mbox{Cov}({\bf Y}, {\bf Z}) =
\mbox{Cov}({\bf X}, {\bf Z}) = - \frac{4}{3}, \]
and so (iii) of Theorem 5 is not satisfied by
these covariances.

It is a somewhat depressing mathematical fact that even for three
random variables with $n$-values and expectations equal to zero, a
separate investigation seems to be needed for each $n$ to find
necessary and sufficient conditions to have a joint probability
distribution compatible with given means, variances and covariances or
correlations. A more general recursive result would be highly
desirable, but seems not to be known. Such results are pertinent to
the study of multi-valued spin phenomena, the discussion of which we
continue after the next theorem.

\paragraph{Bell's original inequality.} We now return to Theorem 5 for
another look at the inequalities (iii), which assume $E({\bf X}) =
E({\bf Y}) = E({\bf Z}) = 0.$ How do these inequalities relate to
Bell's well-known inequality \cite{Bell64}, written in terms of the
bivariate product moments,

\begin{equation}
 1 + E ({\bf YZ} ) \geq \mid E ( {\bf XY}) - E( {\bf
XZ}) \mid ? 
\end{equation}
Bell's inequality is in fact neither necessary nor sufficient
for the existence of a joint probability distribution of the random
variables ${\bf X,Y}$ and ${\bf Z}$ with values ${\pm 1}$ and
expectations equal to zero. That it is not sufficient is easily seen
from letting all three covariances equal $- \frac{1}{2}$. Then the inequality
is satisfied, for 
\[ 1 - \frac{1}{2} \geq \mid -\frac{1}{2} - (-\frac{1}{2}) \mid \]i.e., \[ \frac{1}{2} \geq 0, \]
but, as is clear from (iii) there can be no joint distribution with
the three covariances equal to $- \frac{1}{2}$, for
\[ - \frac{1}{2} + - \frac{1}{2} + -\frac{1}{2} < -1.\]

Secondly, Bell's inequality is not necessary.  Let $E({\bf XY})
= \frac{1}{2}$, $E({\bf XZ}) = - \frac{1}{2}$, and $E({\bf YZ}) = -
\frac{1}{2}$, then $(4)$ is violated, because
\[ 1 - \frac{1}{2} < \mid \frac{1}{2} - (- \frac{1}{2}) \mid, \]
but (iii) is satisfied, and so there is a joint distribution: 
\[ -1 \leq \frac{1}{2} -\frac{1}{2} -\frac{1}{2} \leq 1 + 2 \mbox{{\rm
Min}} (\frac{1}{2}, -\frac{1}{2}, -\frac{1}{2}),\] i.e., 
\[ -1 \leq - \frac{1}{2} \leq 0. \]

Bell derived his inequality for certain cases satisfied by a local
hidden-variable theory, but violated by the quantum mechanical
covariance equal to $- \cos \theta_{ij}.$ In particular, let
$\theta_{{\bf XY}} = 30^{o}, \theta_{{\bf XZ}} = 60^{\circ},
\theta_{{\bf YZ}} = 30^{o}$, so, geometrically ${\bf Y}$ bisects
${\bf X}$ and ${\bf Z}$. Then 
\[ \mid - \frac{1}{2} - \left( - \frac{\sqrt{3}}{2} \right) \mid > 1 -
\frac{\sqrt{3}}{2}.\]

\vskip 0.2 in
{\bf Bell's Inequalities in the CHSH form.}
The next theorem states two conditions equivalent to Bell's
Inequalities for random variables with just two values. This form is
due to Clauser et al., \cite{Clauser-78}. The equivalence of (ii) and
(iii) was proved by Fine \cite{Fine}. 
\begin{theorem}[Bell's Inequalities]
Let $n$ random variables be given satisfying the locality hypothesis
of Theorem 4. Let $n=4$, the number of random variables, let each
${\bf X}_i$ be discrete with values $\pm 1$, let the symmetry
condition $E({\bf X}_i)=0$, $i=1,\ldots ,4$ be satisfied, let ${\bf
X}_1={\bf A}$, ${\bf X}_2={\bf A}'$, ${\bf X}_3={\bf B}$, ${\bf
X}_4={\bf B}'$, with the covariances $E({\bf AB})$, $E({\bf AB}')$,
$E({\bf A}'{\bf B})$ and $E({\bf A}'{\bf B}')$ given. Then the
following three conditions are equivalent.
\begin{description}

\item[(i)] There is a hidden variable {\boldmath $\lambda$} satisfying
Locality Condition II and equation (a) of the Second-Order
Factorization Condition holds.

\item[(ii)] There is a joint probability distribution of the random
variables ${\bf A}$, ${\bf A}'$, ${\bf B}$ and ${\bf B}'$ compatible
with the given means and covariances.

\item[(iii)] The random variables ${\bf A}$, ${\bf A}'$, ${\bf B}$ and
${\bf B}'$ satisfy Bell's inequalities in the CHSH form
\[
-2 \leq E({\bf AB}) + E({\bf AB}') + E({\bf A}'{\bf B}) - E({\bf
A}'{\bf B}') \leq 2
\]
\[
-2 \leq E({\bf AB}) + E({\bf AB}') - E({\bf A}'{\bf B}) + E({\bf
A}'{\bf B}') \leq 2
\]
\[
-2 \leq E({\bf AB}) - E({\bf AB}') + E({\bf A}'{\bf B}) + E({\bf
A}'{\bf B}') \leq 2
\]
\[
-2 \leq - E({\bf AB}) + E({\bf AB}') + E({\bf A}'{\bf B}) + E({\bf
A}'{\bf B}') \leq 2
\]
\end{description}
\end{theorem}
It is worth emphasizing that in contrast to Bell's original inequality
(4), the CHSH inequalities with four random variables give necessary
and sufficient conditions for the existence of a joint probability
distribution.

It will now be shown that the CHSH inequalities 
remain valid for three-valued random variables, (spin-1 particles).
Consider a spin-1 particle with the 3 state observables, $A(a,\lambda)
=+1,0,-1$, $B(b,\lambda)=+1,0,-1$. $\lambda$ is a hidden variable
having a normalized probability density, $\rho (\lambda )$. The
expectation of these observables is defined as,
\begin{eqnarray}
E(a,b) &=& \int AB\rho (\lambda )d\lambda. \nonumber
\end{eqnarray}
We have suppressed the variable dependence on $A$ and $B$ for clarity.
(Note that in this discussion we follow the notation of physicists,
especially as used by Bell, rather than the standard notation of
mathematical statistics for expectations, including covariances.)
Consider the following difference,
\begin{eqnarray}
|E(a,b)-E(a,b')| &=& |\int A[B-B'] \rho (\lambda )d\lambda|. \nonumber
\end{eqnarray}
Since the density $\rho >0$ and $|A|=1,0$ we have the following
inequality,
\begin{eqnarray}
|E(a,b)-E(a,b')| &\leq& \int |A[B-B']| \rho (\lambda )d\lambda,
\nonumber \\
&\leq& \int |[B-B']| \rho (\lambda )d\lambda. \nonumber
\end{eqnarray}
Similarly we have the following inequality,
\begin{eqnarray}
|E(a',b)+E(a',b')| &=& |\int A'[B+B'] \rho (\lambda )d\lambda|,
\nonumber \\
&\leq& \int |[B+B']| \rho (\lambda )d\lambda. \nonumber
\end{eqnarray}
Adding the two expressions we arrive at the following inequality,
\begin{eqnarray*}
|E(a,b)-E(a,b')|+|E(a',b)+E(a',b')| &\!\!=\!\!\!& \int [|B-B'|+|B+B'|]
\rho (\lambda )d\lambda. \\
\end{eqnarray*}
The term in square brackets is equal to 2 in all cases except when $B$
and $B'$ are both equal to zero, in which case the right-hand side vanishes. 
With this and the normalization condition for the hidden variable
density we have the same inequality as the spin-${1\over 2}$ CHSH inequality,
\begin{eqnarray}
|E(a,b)-E(a,b')|+|E(a',b)+E(a',b')| &\leq& 2. \nonumber
\end{eqnarray}
Note that we could create a stronger inequality by adding the function
$2(|E(a,b)|-1)(|E(a,b')|-1)$ to the left-hand side. 

\paragraph{Higher Spin Cases.} For higher spins we can proceed analogously and derive the following
inequality which must be satisfied for spin $j$ particles,
\begin{eqnarray}
|E(a,b)-E(a,b')|+|E(a',b)+E(a',b')| &\leq& 2j. \nonumber
\end{eqnarray} 
If we define normalized observables, $A(a,\lambda)\over j$ the 
original CHSH inequality will need to be satisfied for local
hidden variable theories, although stronger inequalities could be
constructed. 

In Peres' work on higher spin particles the observable is defined
by a mapping from the, $2j+1$-state, $J_z$ operator to a two-state
operator \cite{Peres}. Under this mapping it was shown that Bell's inequality 
is violated for certain parameter settings of the detectors.

The mapping from many values to $\pm1$, as used by Peres and others is
justified probabilistically by the following theorem, which provides a
way of avoiding deriving separate inequalities for each of the higher
spin cases $(n>2)$.

\begin{theorem} Let ${\bf X}_{1}, \ldots , {\bf X}_{n}$ be $n$ random
variables with joint probability distribution $F(x_{1}, \ldots,
x_{n}).$ Let $f_{1}, \ldots , f_{k}$ be finite-valued measurable
functions of the random variables ${\bf X}_{1}, \ldots, {\bf X}_{n}$,
with $y_{1} = f_{1}(x_{1}, \ldots, x_{n}), \ldots, y_{k} =
f_{k}(x_{1}, \ldots, x_{n}).$ Then there is a function $G(y_{1},
\ldots , y_{k})$, unique up to sets of measure zero, that determines
the joint probability distribution of the random variables ${\bf
Y}_{1}, \ldots, {\bf Y}_{k}$ that are functions of ${\bf X}_{1},
\ldots, {\bf X}_{n}.$ \end{theorem}

\noindent{\it Idea of the proof}: We only sketch the proof for a
simple finite case to avoid technical details, for the underlying idea
is very intuitive.

Let ${\bf X}, {\bf Y}$ and ${\bf Z}$ be $\pm 1$ random variables with
a joint distribution. Let ${\bf A}$ and ${\bf B}$ be random variables
that are functions of ${\bf X}$, ${\bf Y}$ and ${\bf Z}$. In
particular, let
\begin{eqnarray*}
{\bf A} = & f({\bf X}, {\bf Y}) = & {\bf X} + {\bf Y} \\
{\bf B} = & f({\bf Y}, {\bf Z}) = & {\bf Y} + {\bf Z}.
\end{eqnarray*} 
Then it is easy to see that range of values of ${\bf A}$ and ${\bf B}$
is $\{-2, 0, 2\}$. More importantly, the joint distribution of ${\bf
A}$ and ${\bf B}$ is easily computed from the joint distribution of ${\bf
X}, {\bf Y} \mbox{ and } {\bf Z}$. Of the nine possible triples of
values for the joint distribution, we show four, the remaining five
are very similar:
\begin{eqnarray*}
P({\bf A}=-2 \:\& \:{\bf B}=-2) & = & P({\bf X}=-1 \:\&\: {\bf Y}=-1 \:\&\: {\bf Z} = -1) \\
P({\bf A}=-2 \:\&\: {\bf B}= 0) & = & P({\bf X}=-1 \:\&\: {\bf Y}=-1 \:\&\: {\bf Z} = 1) \\
P({\bf A}=-2 \:\&\: {\bf B}= 2) & = & 0 \\
P({\bf A}= 0 \:\&\: {\bf B}= 0) & = & P(({\bf X}=-1 \:\&\: {\bf Y}= 1 \:\&\: {\bf Z} =
-1) \mbox { or }\\
                          &   & \;\;\:\:({\bf X}= 1 \:\&\: {\bf Y}=-1
\:\&\: {\bf Z} = 1))
\end{eqnarray*}

The following partial converse of Theorem 7 is really what is implicit
in the reduction of higher spin cases to just two values, rather than
Theorem 7 itself. For simplicity of formulation we restrict the
statement of the theorem to four random variables, using the familiar
notation of Theorem 6, and also restrict the functions to functions
of a single random variable, with the additional constraint that the
functions have only the values $\pm 1$.

\begin{theorem} 
Let ${\bf A}, {\bf B}, {\bf A}^{'}, {\bf B}^{'}$ be
random variables with means, variances and covariances given, but with
no assumption of a joint distribution. Let $f_{{\bf A}}, f_{{\bf B}},
f_{{\bf A}^{'}}, f_{{\bf B}^{'}}$ be finite-valued measurable
functions of the respective random variables and let the functions
have only the values $\pm 1$. If there is no joint distribution of
$f_{{\bf A}}({\bf A}), f_{{\bf B}}({\bf B}), f_{{\bf A}^{'}}({\bf
A}^{'}) \mbox{ and } f_{{\bf B}^{'}}({\bf B}^{'})$ compatible with the means,
variances and covariances of the functional random variables, then
there is no joint distribution of ${\bf A}, {\bf B}, {\bf A}^{'}, {\bf
B}^{'}$ compatible with the given means, variances and covariances.
\end{theorem}

\paragraph{GHZ Probabilistic Theorem.} Changing the focus, we now
consider an abstract version of the GHZ gedanken experiment.  All
arguments known to us, in particular GHZ's \cite{GHZ} own argument,
the more extended one in \cite{GHSZ} and Mermin's \cite{Mermin1}
proceed by assuming the existence of a deterministic hidden variable
and then deriving a contradiction. It follows immediately from Theorem
1 that the nonexistence of a hidden variable is equivalent to the
nonexistence of a joint probability distribution for the given
observable random variables.  The next theorem states this purely
probabilistic GHZ result, and, more importantly, the proof is purely
in terms of the observables, with no consideration of possible hidden
variables.

\begin{theorem}[Abstract GHZ version].  Let ${\bf A}_{\varphi_{1}},
{\bf B}_{\varphi_{2}}, {\bf C}_{\varphi_{3}}, {\bf D}_{\varphi_{4}}$
be an infinite family of $\pm 1$ random variables, with $\varphi_{i}$
a periodic angle or phase, $0 \leq \varphi_{i} \leq 2 \pi$, and let
the following condition hold:
\begin{equation}
E({\bf A}_{\varphi_1} {\bf
B}_{\varphi_2} {\bf C}_{\varphi_3} {\bf D}_{\varphi_4}) = -
\cos ( \varphi_1 + \varphi_2 - \varphi_3 - \varphi_4) \label{ghzE}
\end{equation}
Then the finite subset of random variables ${\bf A}_{0}, {\bf B}_{0}, {\bf
C}_{0}, {\bf D}_{0}, {\bf A}_{\pi} {\bf A}_{\frac{\pi}{2}}, {\bf
C}_{\frac{\pi}{2}}, {\bf D}_{\frac{\pi}{2}}$ does not have a joint
probability distribution.
\end{theorem}
\noindent{\it Proof}: We note first, as an immediate consequence of (\ref{ghzE}), 
\begin{description}

\item[(i)] if $\varphi_{1} + \varphi_{2} - \varphi_{3} - \varphi_{4} =
0$ then $E({\bf A}_{\varphi_{1}} {\bf B}_{\varphi_{2}} {\bf
C}_{\varphi_{3}} {\bf D}_{\varphi_{4}}) = -1$,

\item[(ii)] if $\varphi_{1} + \varphi_{2} - \varphi_{3} - \varphi_{4}
= \pi$ then $E({\bf A}_{\varphi_{1}} {\bf B}_{\varphi_{2}} {\bf
C}_{\varphi_{3}} {\bf D}_{\varphi_{4}}) = 1$.
\end{description} The proof proceeds by deriving a contradiction from
the supposition of the existence of a joint probability distribution.
Because conditional probabilities are used repeatedly, we must check
the given condition in each such probability has positive probability.
Let $s_{i}, i = 1, \ldots, 4$ be $+1$ or $-1$. One of the $16$
products of the four signs must have positive probability, in the
sense that 
\begin{equation}
P({\bf A}_{0} = s_{1}, {\bf B}_{0} = s_{2}, {\bf C}_{0} = s_{3}, {\bf
D}_{0} = s_{4}) > 0
\end{equation}
(We do not need to know whether each $s_{i}$ is $+1$ or $-1$.) Then
since the angles sum to $0$, the product
\begin{equation}
s_1s_2s_3s_4 = -1.
\end{equation}
We also can infer at once from (\ref{ghzE}) and (ii)
\begin{equation}
P({\bf A}_{\pi} = s_{2}s_{3}s_{4} \mid {\bf B}_{0} = s_{2}, {\bf C}_{0}
= s_{3}, {\bf D}_{0} = s_{4}) = 1, \label{Api}
\end{equation}
since (\ref{ghzE}) ensures that the condition in (\ref{Api}) has positive
probability. Using (i) now, by a similar argument
\begin{equation}
P({\bf A}_{0}{\bf C}_{0} = -s_2s_4 \mid {\bf B}_{0} = s_{2}, {\bf
D}_{0} = s_{4}) = 1, \label{A0C0}
\end{equation}
and from $(\ref{ghzE})$ and familiar facts about probability-1
propositions (see Lemma $1$ of the Appendix),
we may add ${\bf C}_{0} = s_{3}$ to the condition (\ref{A0C0}) to obtain
\begin{equation}
P({\bf A}_0 {\bf C}_0 = -s_2s_4 \mid {\bf B}_0 = s_2, {\bf
C}_0 {\bf D}_0 = s_3s_4) = 1. \label{AOCO2}
\end{equation}
Using (i) and (\ref{ghzE}) again
\begin{equation}
P({\bf A}_{\frac{\pi}{2}}{\bf C}_{\frac{\pi}{2}} = -s_2s_4 \mid {\bf
B}_0 = s_2, {\bf C}_0 {\bf D}_0 = s_3s_4) = 1 \label{Api2}
\end{equation}
And so, using Lemma $2$ of the Appendix and (\ref{AOCO2}) and (\ref{Api2}), we infer
\begin{equation}
P({\bf A}_{0} {\bf C}_{0} = {\bf A}_{\frac{\pi}{2}}{\bf
C}_{\frac{\pi}{2}} \mid {\bf B}_{0} = s_{2}, {\bf C}_{0} {\bf D}_{0} =
s_{3}s_{4}) = 1.\label{ACeqAC}
\end{equation}
By an argument just like that of (\ref{A0C0}) - (\ref{ACeqAC}), we
also infer
\begin{equation}
P({\bf A}_0{\bf D}_0 = {\bf A}_{\frac{\pi}{2}} {\bf
D}_{\frac{\pi}{2}} \mid {\bf B}_{0} = s_2, {\bf C}_0 {\bf D}_0 =
s_3s_4) = 1 \label{ADeqAD}
\end{equation}
Dividing the equation of (\ref{ACeqAC}) by that of (\ref{ADeqAD}), we get
\begin{equation}
P\left(\frac{{\bf C}_{0}}{{\bf D}_{0}} = \frac{{\bf C}_{\frac{\pi}{2}}}{{\bf
D}_{\frac{\pi}{2}}} \mid {\bf B}_{0} = s_{2}, {\bf C}_{0} {\bf D}_{0} =
s_{3}s_{4}\right) = 1, \label{C/DeqC/D}
\end{equation}
and since the random variables have only values $+1$ and $-1$, we may
rewrite (\ref{C/DeqC/D}) as:
\begin{equation}
P({\bf C}_{0} {\bf D}_{0} = {\bf C}_{\frac{\pi}{2}} {\bf
D}_{\frac{\pi}{2}} \mid {\bf B}_0 = s_2, {\bf C}_0 {\bf D}_0 =
s_3s_4) = 1 \label{CDeqCD}
\end{equation}

From (\ref{CDeqCD}) and Lemma $3$ of the Appendix we get
\begin{equation}
P( {\bf C}_{\frac{\pi}{2}} {\bf D}_{\frac{\pi}{2}} = s_{3}s_{4} \mid
{\bf B}_{0} = s_{2}, {\bf C}_{0} {\bf D}_{0} = s_{3}s_{4}) = 1, \label{Cp2Dp2}
\end{equation}
and so immediately we may infer from (\ref{ghzE}) and (\ref{Cp2Dp2})
\begin{equation}
P( {\bf B}_{0}= s_{2}, {\bf C}_{\frac{\pi}{2}} {\bf D}_{\frac{\pi}{2}} =
s_3s_4) > 0, \label{BCDgr0}
\end{equation}
Then from $(i)$ and (\ref{BCDgr0})
\begin{equation}
P( {\bf A}_{\pi} = - s_2s_3s_4 \mid {\bf B}_{0} = s_{2}, {\bf
C}_{\frac{\pi}{2}} {\bf D}_{\frac{\pi}{2}} = s_{3}s_{4}) = 1, \label{Apeq234}
\end{equation}
and finally from (\ref{CDeqCD}) and (\ref{Apeq234}) and Lemma $5$ of the Appendix
\begin{equation}
P( {\bf A}_{\pi} = - s_{2}s_{3}s_{4} \mid {\bf B}_{0} = s_{2}, {\bf
C}_{0} {\bf D}_{0} = s_{3}s_{4}) = 1.\label{Ape234b}
\end{equation}

Obviously, (\ref{Api}) and (\ref{Ape234b}) together yield the desired contradiction.

\paragraph{Gaussian random variables.} A fundamental second-order theorem about finite sequences of
continuous random variables is the following:

\begin{theorem} Let $n$ continous random variables be given, let their
means, variances and covariances all exist and be finite, with all the
variances nonzero.  Then a necessary and sufficient condition that a
joint Gaussian probability distribution of the $n$ random variables exists,
compatible with the given means, variances and covariances, is that
the eigenvalues of the correlation matrix be nonnegative.
\end{theorem} A thorough discussion and proof of this theorem can be
found in Lo\`eve \cite{Loeve}. It is important to note that the
hypothesis of this theorem is that each pair of the random variables
has enough postulated for there to exist a unique bivariate Gaussian
distribution with the given pair of means and variances and the
covariance of the pair. Moreover, if, as required for a joint
distribution of all $n$ variables, the eigenvalues of the correlation
matrix are all nonnegative, then there is a unique Gaussian joint
distribution of the $n$ random variables. 

We formulate the next theorem to include cases like Bell's
inequalities when not all the correlations or covariances are given.
\begin{theorem}
Let $n$ continuous random variables be given such that they satisfy
the locality hypothesis of Theorem 4, let their means and variances
exist and be finite, with all the variances nonzero, and let $m \leq
n(n-1)/2$ covariances be given and be finite. Then the following two
conditions are equivalent.
\begin{description}

\item[(i)] There is a joint Gaussian probability distribution of the
$n$ random variables compatible with the given means, variances and
covariances.

\item[(ii)] Given the $m \leq n(n-1)/2$ covariances, there are real
numbers that may be assigned to the missing correlations so that the
completed correlation matrix has eigenvalues that are all nonnegative.

\end{description}

Moreover, $(i)$ or $(ii)$ implies that there is a hidden variable 
{\boldmath $\lambda$} satisfying
Locality Condition II and the Second-Order Factorization Condition.

\end{theorem}
The proof of Theorem 11 follows directly from Theorem 10.

Using Theorem 10, we can also derive a nonlinear inequality necessary
and sufficient for three Gaussian random variables to have a joint
distribution.  In the statement of the theorem $\rho ({\bf X,
Y})$ is the correlation of {\boldmath $X$} and {\boldmath $Y$}.

\begin{theorem}
Let ${\bf X, Y}$ and ${\bf Z}$ be three Gaussian random variables
whose means, variances and correlations are given, and whose variances
are nonzero.  Then there exists a joint Gaussian distribution of ${\bf
X, Y}$ and ${\bf Z}$ (necessarily unique) compatible with the given
means, variances and correlations if and only if 
\[
\rho ({\bf X,Y})^{2}
+ \rho ({\bf X,Z})^{2} + \rho ({\bf Y,Z})^{2} \leq 2 \rho ({\bf X,Y})
\rho ({\bf Y,Z}) \rho ({\bf X,Z}) + 1.
\]
\end{theorem}

The proof comes directly from the determinant of the correlation
matrix. For a matrix to be non-negative definite the determinant
of the entire matrix
and all principal minors must be greater than or equal to zero,

\begin{eqnarray}
Det\left(\matrix{1 & \rho({\bf X,Y}) &\rho({\bf X,Z}) \cr
\rho ({\bf X,Y}) & 1 & \rho({\bf Y,Z}) \cr
\rho ({\bf X,Z}) & \rho ({\bf Y,Z}) & 1 \cr}\right) & \geq & 0. \nonumber \\
\end{eqnarray}
Including the conditions for the minors we have,
\begin{eqnarray}
\rho({\bf X,Y})^2+\rho({\bf X,Z})^2+\rho({\bf Y,Z})^2
-2\rho({\bf X,Y})\rho({\bf X,Z})\rho({\bf Y,Z}) & \leq & 1
\nonumber \\
\rho({\bf X,Y})^2 &\leq & 1 \nonumber \\
\rho({\bf Y,Z})^2 &\leq & 1 \nonumber \\
\rho({\bf X,Z})^2 &\leq & 1.
\end{eqnarray}
The last three inequalities are automatically satisfied since the
correlations are bounded by $\pm 1$.

\paragraph{Simultaneous observations and joint distributions.}
When observations are simultaneous and the environment is stable and
stationary, so that with repeated simultaneous observations
satisfactory frequency data can be obtained, then there exists a joint
distribution of all of the random variables representing the
simultaneous observations.  Note what we can then conclude from the
above: in all such cases there must be, therefore, a
factorizing hidden variable because of the existence of the joint
probability distribution. From this consideration alone, it follows
that any of the quantum mechanical examples that violate Bell's
inequalities or other criteria for hidden variables must be such that
not all the observations in question can be made simultaneously. The
extension of this criterion of simultaneity to a satisfactory
relativistic criterion is straightforward.

\section{Appendix}

We prove here several elementary lemmas about probability-$1$
statements used in the proof of Theorem $9$.

\begin{lemma}
If $P(A \mid B) = 1$ and $P(BC) > 0$ then $P(A \mid BC) =
1.$ 
\end{lemma}
{\it Proof}.  Suppose, by way of contradiction, that
\begin{equation}
P(A \mid BC) < 1. \label{lemma1-1}
\end{equation}
Now from (\ref{lemma1-1})
and the definition of conditional probability, we have at once
\begin{equation}
P(ABC) < P(BC).\label{lemma1-2}
\end{equation}
Adding $P(AB \overline{C})$ to both sides of (\ref{lemma1-2}) and 
simplifying we have
\begin{equation}
P(AB) < P(BC) + P(AB \overline{C}).\label{lemma1-3}
\end{equation}

We now take conditional probabilities with respect to $B$, and
divide both sides of (\ref{lemma1-3}) by $P(B)$, for by the hypothesis of the
lemma, $P(B) > 0$, and thus we obtain
\[
P(A \mid B) < P(C \mid B) + P(A \overline{C} \mid B),
\]
but
\[
P(C \mid B ) + P(A \overline{C} \mid B) \leq 1
\]
and by the hypothesis of the lemma
\[
P(A \mid B) = 1,
\]
whence we have derived the absurdity that $1 < 1.$ Thus the lemma is established.

\begin{lemma}
Let ${\bf X}$ and ${\bf Y}$ be two random variables with a
joint distribution, and let
\begin{description}
\item[(i)] $P(A)>0,$

\item[(ii)] $P({\bf X}=c \mid A) = 1$,

\item[(iii)] $P({\bf Y} = c \mid A) = 1$.
\end{description}

Then
\end{lemma}
\[
P({\bf X} = {\bf Y} \mid A) = 1.
\]
{\it Proof}. Let
\begin{eqnarray*} 
B & = & \{ \omega : {\bf X} (\omega) = c \} \\
C & = & \{ \omega : {\bf Y} (\omega) = c \} \\
D & = & \{ \omega : {\bf X} (\omega) = {\bf Y}(\omega) \}
\end{eqnarray*}
Suppose by way of contradiction that
\[
P(D \mid A ) < 1.
\]
Then
\[
P(\{ \omega : {\bf X} (\omega) \neq {\bf Y} (\omega) \} \mid A) > 0
\]
And so
\[
P( \{ \omega : X (\omega) \neq c \mbox{ or }Y (\omega) \neq c \} \mid A) > 0.
\]
Without loss of generality, let
\[
P( \{ \omega : {\bf X} (\omega) \neq c \} \mid A) > 0.
\]
Then
\[
P( \overline{B} \mid A) > 0,
\]
and this contradicts (ii).

We also need a sort of converse of Lemma $2$.

\begin{lemma}
If $P(A \, \& \, {\bf X} = c) > 0$ and
$P({\bf X} = {\bf Y} \mid A \, \& \, {\bf X} = c) = 1$ then
\[
P({\bf Y} = c \mid A \, \& \, {\bf X} = c) = 1.
\]
\end{lemma}
{\it Proof}. By hypothesis
\[
P({\bf X} = {\bf Y} \, \& \, A \, \& \, {\bf X} = c ) = P(A \, \& \, {\bf X} = c).
\]
Consider now the left-hand side:
\begin{eqnarray*}
\{\omega : X(\omega ) = Y(\omega ) \} \, \& \, \{ \omega : X (\omega) = c \} & =
& \{ \omega : {\bf X} (\omega) = c \, \& \, {\bf Y} (\omega) = c \} \\
& = & \{ \omega : {\bf X} (\omega) = c \} \cap \{ \omega : {\bf Y}
(\omega) = c \},
\end{eqnarray*}
and so
\[
P({\bf X} = {\bf Y} \, \& \, A \, \& \, {\bf X} = c) = P({\bf Y} = c
\, \& \, A \, \& \, {\bf
X} = c),
\]
and thus,
\[
P({\bf Y} = c \, \& \, A \, \& \, {\bf X} = c) = P(A \, \& \, {\bf X} = c),
\]
whence
\[
P({\bf Y} = c \mid A \, \& \, {\bf X} = c) = 1.
\]

We can also prove a kind of transitivity for conditional probabilities
that are $1$.

\begin{lemma}
If $P(B) > 0, P(C) > 0, P(A \mid B) = 1$ and $P(B \mid C) = 1$, then
$P(A \mid C) = 1$.
\end{lemma}
{\it Proof}. By hypothesis and Lemma $1$
\[
P(A \mid BC) = 1,
\]
so
\[
P(ABC) = P(BC)
\]
but by hypothesis
\[
P(BC)= P(C),
\]
so
\[
P(ABC) = P(C),
\]

and thus
\[
P(AB \mid C)= 1,
\]
whence
\[
P(A \mid C) = 1.
\]

Finally, we also use the following,

\begin{lemma}
If $P(A \, \& \, {\bf Y} = d) > 0, P(A \, \& \, {\bf Z} = d) > 0$, and
\[
(i) P({\bf X} = c \mid A \, \& \, {\bf Y} = d) = 1,
\]
\[
(ii) P({\bf Z} = {\bf Y} \mid A \, \& \, {\bf Z} = d) = 1,
\]
then
\end{lemma}
\[
P({\bf X} = c \mid A \, \& \, {\bf Z} = d) = 1.
\]
{\it Proof}. By Lemma $3$ and (ii)
\[
P(A \, \& \, {\bf Y} = d \mid A \, \& \, {\bf Z} = d) = 1
\]
So, by transitivity (Lemma $4$) \& (i)
\[
P({\bf X} = c \mid A  \, \& \, {\bf Z} = d ) = 1
\]

\vspace{1em}

\paragraph{Acknowledgments.} We are grateful to Martin Jones for
useful comments and criticisms of an earlier draft. J.~A.~B.
acknowledges financial support from FAPEMIG (Minas Gerais State
Support Agency) and support from the Laboratory for Experimental
Cosmology and HEP (Lafex) of the Brazilian Centre for Physical
Research (CBPF).


\begin{thebibliography}{99} 

\bibitem{Bell64} BELL J. S., {\em Physics}, {\bf 1}, (1964) 195.

\bibitem{Bell66} BELL J.\ S.\ , {\em Rev. Mod. Phys.} {\bf 38}, (1966) 447.

\bibitem{Clauser-78} CLAUSER J.\ F.\ and SHIMONY A.\ ,
{\em Rep.\ Prog.\ Phys.} {\bf 41}, (1978) 1881.

\bibitem{Fine} FINE A.\ , {\em Phys.\ Rev.\ Lett.} {\bf 48}, (1982) 291.

\bibitem{GHZ} GREENBERGER D.\ M.\ , HORNE M.\ A.\ and ZEILINGER A.\,
``Going beyond Bell's theorem'', in {\em Bell's Theorem, Quantum
Theory, and Conceptions of the Universe}, edited by M.\ KAFATOS
(Kluwer Academic, Dordrecht, The Netherlands) 1989, 69--72.

\bibitem{GHSZ} GREENBERGER D. M., HORNE M. A., SHIMONY A. and
ZEILINGER A., {\em Amer.\ J.\ Phys.} {\bf 58}, (1990) 1131.

\bibitem{Holland} HOLLAND P.\ W.\ and ROSENBAUM T.\ R.\, {\em Ann. Statist.} 
{\bf 14}, (1986) 1523.

\bibitem{Loeve} LO\`EVE M.\, {\em Probability Theory II,} 4th edition.
(Springer Verlag, New York) 1978.

\bibitem{Mermin1} MERMIN N.\ D.\, {\em Am.\ J.\ Phys.} {\bf 58}, (1990),
731.

\bibitem{Peres} PERES A.\, {\em Phys.\ Rev.\ A,} {\bf 46}, (1992), 4413.

\bibitem{Suppes-76} SUPPES P.\ and ZANOTTI M.\, in {\em Logic and
Probability in Quantum Mechanics}, edited by P.\ SUPPES (Reidel,
Dordrecht) 1976, 445.

\bibitem{SuppesZanotti80} SUPPES P.\ and ZANOTTI M.\, ``A new proof of
the impossibility of hidden variables using the principles of
exchangeability and identity of conditional distributions'', in {\em
Studies in the Foundations of Quantum Mechanics}, edited by P.\ SUPPES
(Philosophy of Science Association, East Lansing, Michigan) 1980, 173-191.

\bibitem{SuppesZanotti81} SUPPES P.\ and ZANOTTI M.\, {\em Synthese} {\bf
48}, (1981) 191.

\end{thebibliography}
\end{document}